\documentclass[journal,10pt]{IEEEtran}

\usepackage{graphicx}
\graphicspath{{figures/}}
\usepackage{tikz}

\usepackage{amsmath,amsfonts,stmaryrd,dsfont}
\usepackage{multirow}
\usepackage[linesnumbered,algoruled]{algorithm2e}

\usepackage{array}
\usepackage{comment}

\usepackage{url}

\begin{document}

\title{Online Spectrogram Inversion for \\Low-Latency Audio Source Separation}
\author{Paul~Magron,
	Tuomas~Virtanen,~\IEEEmembership{Senior Member,~IEEE}
	\thanks{P. Magron is with IRIT, Universit\'e de Toulouse, CNRS, France (paul.magron@irit.fr). T. Virtanen is with the Audio Research Group, Tampere University, Finland (tuomas.virtanen@tuni.fi). The work of P. Magron was conducted while he was with Tampere University and supported by the Academy of Finland, project no.\ 290190.}
}


\maketitle

\begin{abstract}
Audio source separation is usually achieved by estimating the short-time Fourier transform (STFT) magnitude of each source, and then applying a spectrogram inversion algorithm to retrieve time-domain signals. In particular, the multiple input spectrogram inversion (MISI) algorithm has been exploited successfully in several recent works. However, this algorithm suffers from two drawbacks, which we address in this paper. First, it has originally been introduced in a heuristic fashion: we propose here a rigorous optimization framework in which MISI is derived, thus proving the convergence of this algorithm. Besides, while MISI operates offline, we propose here an online version of MISI called oMISI, which is suitable for low-latency source separation, an important requirement for e.g., hearing aids applications. oMISI also allows one to use alternative phase initialization schemes exploiting the temporal structure of audio signals. Experiments conducted on a speech separation task show that oMISI performs as well as its offline counterpart, thus demonstrating its potential for real-time source separation.
\end{abstract}

\begin{IEEEkeywords}
Audio source separation, low-latency, online spectrogram inversion, phase recovery, sinusoidal modeling.
\end{IEEEkeywords}

\IEEEpeerreviewmaketitle

\section{Introduction}

Audio source separation~\cite{Comon2010} consists in extracting the underlying \textit{sources} that add up to form an observable audio \textit{mixture}. This task finds applications in many areas such as speech enhancement and recognition~\cite{Barker2018}, musical signal processing~\cite{Cano2019} and hearing aid devices~\cite{Wang2008}. In particular, hearing aids require a very low processing latency, as significant discomfort can be experienced by listeners for delays exceeding 20 ms~\cite{Agnew2000}.  State-of-the-art approaches for source separation consist in using a deep neural network (DNN) to estimate a nonnegative mask that is applied to a time-frequency (TF) representation of the audio mixture, such as the short-time Fourier transform (STFT)~\cite{Wang2018c}. Recent works such as~\cite{Luo2018,Luo2019} operate in the time domain directly, but TF approaches remain interesting since they make it possible to better exploit the structure of sound.

Applying a nonnegative mask to the input STFT results in assigning the mixture's phase to each isolated source. Even though this yields somewhat satisfactory results in practice, it has been pointed out~\cite{Magron2015} that when sources overlap in the TF domain, using the mixture's phase induces residual interference and artifacts in the estimates. With the advent of deep learning, magnitudes can nowadays be estimated with a high accuracy, which highlights the need for advanced phase recovery~\cite{Gerkmann2015, Mowlaee2016}, a problem hereafter termed \emph{spectrogram inversion}. Consequently, several recent works have focused on phase recovery in DNN-based source separation, whether phase recovery algorithms are applied as a post-processing~\cite{Wang2018,Magron2018b} or integrated within end-to-end systems for time-domain separation~\cite{Wang2018a,Wichern2018,Wang2019a,Wisdom2019,LeRoux2019a}.

Among the variety of phase recovery techniques, the multiple input spectrogram inversion (MISI) algorithm~\cite{Gunawan2010} is particularly popular. This iterative procedure consists in retrieving time-domain sources from their STFT magnitudes while respecting a mixing constraint: the estimates must add up to the original mixture. This algorithm has shown promising for audio source separation when combined with DNNs~\cite{Wang2018,Wang2018a}. However, the MISI algorithm has been introduced in a heuristic fashion, therefore there is currently no proof that it converges. Besides, while several recent works addressed the problem of low-latency magnitude estimation~\cite{Naithani2017,Aihara2019,Wang2019}, the MISI algorithm operates offline, as it computes the whole STFT and its inverse at each iteration. This makes it impracticable for real-time applications such as hearing aids.


In this paper, we investigate and overcome the drawbacks of the MISI algorithm. First, we propose a rigorous optimization framework for spectrogram inversion. Using the auxiliary function method we derive a procedure that is equivalent to the MISI algorithm. To the best of our knowledge, this is the first proof of convergence of MISI. Second, we propose an online adaptation of MISI that is suitable for low-latency applications. This adaptation is based on approximating the STFT and its inverse in a causal fashion by only accounting for the past context and for an arbitrarily small number of future frames. It also allows us to exploit the temporal structure of the phase to use alternative phase initialization schemes. Here, we propose to use a sinusoidal phase~\cite{Mowlaee2015} instead of the mixture's phase as initial estimate. We experimentally demonstrate the potential of this technique for a speech separation task.

The rest of this paper is structured as follows. Section~\ref{sec:notations} presents some mathematical notations. The MISI algorithm is derived in Section~\ref{sec:algo} and adapted online in Section~\ref{sec:online}. Section~\ref{sec:exp} presents the experimental results, and Section~\ref{sec:conclusion} concludes the paper.

\section{Mathematical notations}
\label{sec:notations}

\begin{itemize}
    \item $\mathbf{A}$ (capital, bold font): matrix.
    \item $\tilde{\mathbf{x}}$ (lower case, bold font, with tilde): time-domain signal, whose $n$-th sample is denoted $\tilde{\mathbf{x}}(n)$.
    \item  $\mathbf{x}$ (lower case, bold font, without tilde): TF domain vector, whose $m$-th entry is denoted $\mathbf{x}(m)$.
    \item $z$ (regular): scalar.
    \item $|.|$, $\angle(.)$: magnitude and complex angle, respectively.
    \item $\mathbf{x}^\mathsf{T}$, $\mathbf{x}^\mathsf{H}$: transpose and Hermitian transpose, respectively.
    \item $\Re(.)$: real part function.
    \item $||.||$: Euclidean norm.
    \item $\odot$, $\frac{.}{.}$: element-wise matrix or vector multiplication, and division, respectively.
\end{itemize}

\section{Algorithm derivation}
\label{sec:algo}

In this section we derive MISI using the auxiliary function method, which proves the convergence of this algorithm.

\subsection{Problem setting}


Let us consider an instantaneous and linear mixing model:
%
\begin{equation}
\tilde{\mathbf{x}} = \sum_{j=1}^J \tilde{\mathbf{s}}_j,
\label{eq:mixmodel}
\end{equation}
where $\tilde{\mathbf{x}} \in \mathbb{R}^{N}$ is the mixture, $\tilde{\mathbf{s}}_j \in \mathbb{R}^{N}$ are the $J$ sources, and $N$ denotes the length of the time-domain signals in samples. The STFTs of the sources are denoted by $\mathbf{s}_j \in \mathbb{C}^M$ with $M=F \times T$, where $F$ and $T$ are the number of frequency channels and time frames respectively, and we have
\begin{equation}
\mathbf{s}_j = \mathbf{A} \tilde{\mathbf{s}}_j,
\label{eq:stft}
\end{equation}
where the matrix $\mathbf{A} \in \mathbb{C}^{M \times N}$ encodes the STFT operation.

Let us assume that some STFT magnitude estimates denoted $\mathbf{v}_j \in \mathbb{R}_+^M $ are available (e.g., estimated beforehand using a DNN). The goal of multiple-input spectrogram inversion is to estimate time domain source signals $\tilde{\mathbf{s}}_j$ given the STFT magnitude estimates $\mathbf{v}_j$. Since these magnitude estimates are usually not equal to the ground truth, one should allow the magnitudes of the estimated sources to deviate from those values. Thus, we consider the following objective function:
\begin{equation}
\psi(\tilde{\mathbf{s}}) = \sum_{j} || |\mathbf{A} \tilde{\mathbf{s}}_j| -  \mathbf{v}_j ||^2,
\label{eq:mag_function_time}
\end{equation}
We treat the mixing constraint~\eqref{eq:mixmodel} as a hard constraint, leading to the following optimization problem:
\begin{equation}
\min_{\tilde{\mathbf{s}}} \psi(\tilde{\mathbf{s}}) \text{ subject to } \sum_j \tilde{\mathbf{s}}_j = \tilde{\mathbf{x}}.
\label{eq:misi_opt_problem}
\end{equation}
Directly addressing the optimization problem~\eqref{eq:misi_opt_problem} is however difficult since $\psi$ is not differentiable with respect to $\tilde{\mathbf{s}}$. Therefore, we propose to use the auxiliary function method~\cite{Fevotte2011}, which consists in finding a function $\psi^+(\tilde{\mathbf{s}}, \mathbf{y})$ such that %
\begin{equation}
\psi(\tilde{\mathbf{s}}) = \min_{\mathbf{y}}  \psi^+(\tilde{\mathbf{s}},\mathbf{y}),
\label{eq:auxiliary_func}
\end{equation}
where $\mathbf{y}$ is a set of auxiliary parameters. It can be shown~\cite{Fevotte2011} that $\psi$ is non-increasing when alternating the minimization of $\psi^+$ with respect to (w.r.t.) $\tilde{\mathbf{s}}$ and $\mathbf{y}$.

\subsection{Auxiliary function}

Let us first introduce a set of auxiliary parameters $\mathbf{y}_j$ such that $|\mathbf{y}_j|=\mathbf{v}_j$ and rewrite~\eqref{eq:mag_function_time} as:
\begin{equation}
\psi(\tilde{\mathbf{s}}) = \sum_{j} || |\mathbf{A} \tilde{\mathbf{s}}_j| - |\mathbf{y}_{j}| ||^2.
\end{equation}
We then use the property
\begin{equation}
\forall (z,z') \in \mathbb{C}^2 \text{, } ||z|-|z'|| \leq |z-z'|,
\label{eq:ineq}
\end{equation}
that arise from the triangle inequality, and where equality holds if and only if $\angle z = \angle z'$. This leads to $\psi(\tilde{\mathbf{s}}) \leq \psi^+(\tilde{\mathbf{s}},\mathbf{y})$ with:

\begin{equation}
\psi^+(\tilde{\mathbf{s}},\mathbf{y}) = \sum_{j} || \mathbf{A} \tilde{\mathbf{s}}_j - \mathbf{y}_j ||^2.
\end{equation}

In order to minimize $\psi^+$ w.r.t. $\mathbf{y}$ under the constraint $|\mathbf{y}_j| = \mathbf{v}_j$, we introduce this constraint using the Lagrange multipliers method. We therefore aim at finding a saddle point for:

\begin{equation}
\psi^+(\tilde{\mathbf{s}}, \mathbf{y}) + \sum_{j,m} \boldsymbol{\lambda}_{j}(m) (|\mathbf{y}_{j}(m)|^2-\mathbf{v}_{j}(m)^2),
\label{eq:aux_psi}
\end{equation}
where $\boldsymbol{\lambda}_{j} \in \mathbb{R}^M$ are the Lagrange multipliers. We set the partial derivative of~\eqref{eq:aux_psi} w.r.t $\mathbf{y}$ at $0$, which leads to:
\begin{equation}
(1+\boldsymbol{\lambda}_{j}(m)) \mathbf{y}_{j}(m) = \mathbf{s}_{j}(m).
\label{eq:deriv}
\end{equation}
Using the constraint $|\mathbf{y}_{j}(m)|=\mathbf{v}_{j}(m)$, we have
\begin{equation}
|1+\boldsymbol{\lambda}_{j}(m)| = \frac{|\mathbf{s}_{j}(m)|}{\mathbf{v}_{j}(m)}.
\label{eq:lambda}
\end{equation}
Finally, injecting~\eqref{eq:lambda} into~\eqref{eq:deriv} leads to the update for $\mathbf{y}_j$:
\begin{equation}
\mathbf{y}_j = \pm \frac{\mathbf{s}_j}{|\mathbf{s}_j|} \odot \mathbf{v}_j,
\label{eq:Y}
\end{equation}
We consider the update that does not modify the phase of $\mathbf{s}_j$ (i.e., with a '$+$' sign in~\eqref{eq:Y}), as it corresponds to the equality case of Eq.~\eqref{eq:ineq}. Under such an update, $\psi(\tilde{\mathbf{s}}) = \psi^+(\tilde{\mathbf{s}},\mathbf{y})$, which shows that $\psi^+$ is an auxiliary function for $\psi$.

\subsection{Including the mixing constraint}
\label{sec:algo_mainup}

Now, let us introduce the hard mixing constraint~\eqref{eq:mixmodel} within the auxiliary function $\psi^+$ by means of the Lagrange multipliers as in~\cite{Magron2018}. This results in finding a saddle point for:
\begin{equation}
\Psi(\tilde{\mathbf{s}},\mathbf{y},\tilde{\boldsymbol{\delta}}) = \sum_{j} || \mathbf{A} \tilde{\mathbf{s}}_j - \mathbf{y}_j ||^2 + 2 \Re \left( \tilde{\boldsymbol{\delta}}^\mathsf{H} ( \sum_j \tilde{\mathbf{s}}_j - \tilde{\mathbf{x}} ) \right),
\end{equation}
where $\tilde{\boldsymbol{\delta}} \in \mathbb{C}^N$ is the vector of Lagrange multipliers. Setting the derivative of $\Psi$ w.r.t. $\tilde{\mathbf{s}}_j$ at $0$ yields:
\begin{equation}
2 \mathbf{A}^\mathsf{H} \mathbf{A} \tilde{\mathbf{s}}_j - 2 \mathbf{A}^\mathsf{H} \mathbf{y}_j + 2 \tilde{\boldsymbol{\delta}} = 0.
\label{eq:s_temp}
\end{equation}
Let us point out that the matrix $\mathbf{A}^\mathsf{H}$ encodes the inverse STFT~\cite{LeRoux2013}. Indeed, one can show~\cite{Yang2008} that if the synthesis window is equal to the analysis window up to a specific normalization constant~\cite{Griffin1984}, the STFT is an Hermitian operator. Assuming such analysis-synthesis windows are used, $\mathbf{A}^\mathsf{H} \mathbf{A}$ is the identity matrix. We define:
\begin{equation}
\tilde{\mathbf{y}}_j = \mathbf{A}^\mathsf{H} \mathbf{y}_j = \text{iSTFT}(\mathbf{y}_j),
\label{eq:y}
\end{equation}
and therefore Eq.~\eqref{eq:s_temp} rewrites:
\begin{equation}
\tilde{\mathbf{s}}_j -\tilde{\mathbf{y}}_j  + \tilde{\boldsymbol{\delta}} = 0.
\label{eq:deriv_delta}
\end{equation}
Summing~\eqref{eq:deriv_delta} over $j$ and using the mixing constraint yields:
\begin{equation}
\tilde{\mathbf{x}} - \sum_j \tilde{\mathbf{y}}_j  + J \tilde{\boldsymbol{\delta}} = 0.
\end{equation}
Finally, solving for $\tilde{\boldsymbol{\delta}}$ and injecting it in Eq.~\eqref{eq:deriv_delta} leads to:
\begin{equation}
\tilde{\mathbf{s}}_j = \tilde{\mathbf{y}}_j  + \frac{1}{J} (\tilde{\mathbf{x}} - \sum_p \tilde{\mathbf{y}}_p).
\label{eq:update_mix}
\end{equation}
Combining the updates given by Eq.~\eqref{eq:stft},~\eqref{eq:Y},~\eqref{eq:y}, and~\eqref{eq:update_mix} yields the MISI algorithm, as introduced in the original paper~\cite{Gunawan2010}; this derivation therefore proves its convergence.

\section{Online MISI}
\label{sec:online}

First, let us reshape the STFTs $\mathbf{s}_j$ onto matrix form as they are usually processed this way: $\mathbf{S}_j \in \mathbb{C}^{F \times T}$. We rewrite the MISI algorithm in the TF domain, as done in~\cite{Wang2018a,Wichern2018}: $\forall j$,
\begin{align}
&\mathbf{Z}_j = \text{STFT}(  \text{iSTFT} (\mathbf{S}_j)), \label{eq:g} \\
&\mathbf{Y}_j = \frac{\mathbf{Z}_j }{|\mathbf{Z}_j |} \odot \mathbf{V}_j, \label{eq:norm} \\
&\mathbf{S}_j = \mathbf{Y}_j  + \frac{1}{J} (\mathbf{X} - \sum_p \mathbf{Y}_p), \label{eq:mixerrdist}
\end{align}

\subsection{Problem setting}
\label{sec:omisi_problem}
It is straightforward to implement~\eqref{eq:norm} and~\eqref{eq:mixerrdist} online, as these are performed bin-wise, but this is not the case of~\eqref{eq:g}. Indeed, the inverse STFT of $\mathbf{S}_j$ is computed through the overlap-add (OLA) procedure as follows: 
\begin{align}
\tilde{\textbf{s}}'_{j,t} &= \text{iDFT}(\textbf{S}_{j,t}) \odot \tilde{\textbf{w}}, \label{eq:idft}\\
\tilde{\textbf{s}}_{j}(n) &= \sum_{t=0}^{T-1} \tilde{\textbf{s}}'_{j,t} (n-tl), \label{eq:ola}
\end{align}
where $\textbf{S}_{j,t}$ is the $t$-th column of $\textbf{S}_j$, iDFT denotes the inverse discrete Fourier transform, $\tilde{\textbf{w}}$ is a window of length $N_w$, and $l$ is the hop size. For notation convenience, we consider that $\tilde{\textbf{s}}'_{j,t} (n-tl)=0$ if $n \notin \{0,...,N_w-1 \}$.


Using OLA~\eqref{eq:ola}, a sample $n$ is reconstructed by accounting for all frames whose iDFT has a temporal support that includes $n$, which is not suitable for online applications.

\subsection{Online implementation}

\begin{figure}[t]
    \centering
    \includegraphics[width=.75\columnwidth]{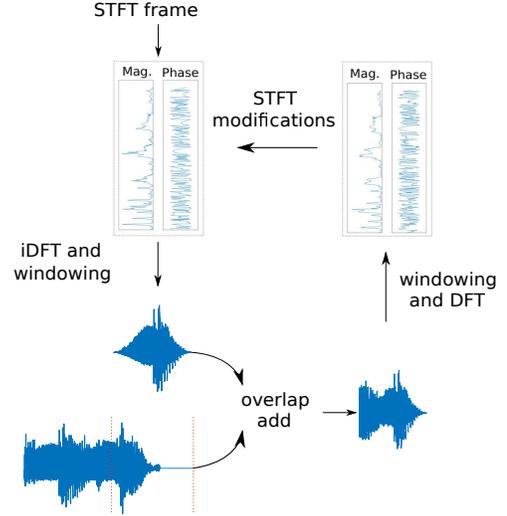}
    \caption{Illustration of the proposed approach in one time frame with $K=0$.}
    \label{fig:approx_g}
\end{figure}

Let us assume that we are currently processing the frame indexed by $t$. We first decompose the OLA procedure~\eqref{eq:ola} as:
\begin{equation}
\tilde{\textbf{s}}_{j}(n) = \underbrace{\sum_{k=0}^{t-1} \tilde{\textbf{s}}'_{j,k} (n-tl)}_{\tilde{\textbf{s}}^{\text{past}}_{j,t}(n)} + \underbrace{\sum_{k=t}^{T-1} \tilde{\textbf{s}}'_{j,k} (n-tl)}_{\tilde{\textbf{s}}^{\text{fut}}_{j,t}(n)},
\end{equation}
where $\tilde{\textbf{s}}^{\text{past}}_{j,t}$ (resp. $\tilde{\textbf{s}}^{\text{fut}}_{j,t}$) contains the contributions of the previous (resp. current and future) frames. Drawing on prior work~\cite{Beauregard2005,Zhu2006,Zhu2007}, we propose hereafter to approximate $\tilde{\textbf{s}}^{\text{fut}}_{j,t}$ by using only the current frame and an arbitrarily small number $K \geq 0$ of future frames:
\begin{equation}
\tilde{\textbf{s}}^{\text{fut}}_{j,t}(n) \approx \sum_{k=t}^{t+K} \tilde{\textbf{s}}'_{j,k} (n-tl).
\label{eq:sfut_approx}
\end{equation}
Even though this approach results in losing the contributions of some of the future time frames involved in the calculation of  $\tilde{\textbf{s}}_{j}(n)$ (\textit{cf}. \ref{sec:omisi_problem}), it still enforces a form of coherence over time by accounting for the overlap with the previous time frames. It also preserves the coherence with the near-future frames which overlap with the current one. With such an approach, the algorithmic latency is reduced to $N_w + Kl$ samples. It is illustrated in Fig.~\ref{fig:approx_g}. This procedure is called oMISI (for ``online MISI") and summarized in Algorithm~\ref{al:oMISI}.

\begin{algorithm}[t]
	\caption{Online MISI}
	\label{al:oMISI}
			\textbf{Inputs}: Mixture $\mathbf{X} \in \mathbb{C}^{F \times T}$, magnitudes $\mathbf{V}_j \in \mathbb{C}^{F \times T}$, number of iterations $N_i$ and future frames $K$. \\

            Initialize the past segments: $\forall j$, $\tilde{\mathbf{s}}_j^{\text{past}} = 0$\\
            
			\For{$t=0$ \KwTo $T-1-K$}{
			Initialize the phase in the new frame $t+K$ (\textit{cf}.~\ref{sec:phase_init})  \\  \label{algo:phase_init}
            
            \For{$iter=1$ \KwTo $N_i$}{
                $\forall j$: Compute $\tilde{\textbf{s}}^{fut}_{j}$ using~\eqref{eq:idft} and~\eqref{eq:sfut_approx} \\
                $\forall j$: $\textbf{Z}_{j} = \text{STFT}( \tilde{\textbf{s}}^{past}_{j}  + \tilde{\textbf{s}}^{fut}_{j})$ \\
                $\forall j$: Update  $\mathbf{S}_{j,t},...,\mathbf{S}_{j,t+K}$ using~\eqref{eq:norm} and~\eqref{eq:mixerrdist}  \\
                }
            
            Compute, $\forall j$, $\tilde{\mathbf{s}}'_{j,t}$ from $\textbf{S}_{j,t}$ using~\eqref{eq:idft} \\
            Update the sources: $\forall j$, $\tilde{\mathbf{s}}_{j}(tl+n) = \tilde{\mathbf{s}}'_{j,t}(n) + \tilde{\mathbf{s}}^{past}_{j}(n)$ for $n<l$\\
            Update the past segments: $\forall j$, $\tilde{\mathbf{s}}^{\text{past}}_{j}(n) = \tilde{\mathbf{s}}'_{j,t}(n+l)$ for $n< N_w-l$ and $0$ otherwise
			}
			
			\textbf{Outputs}: $\tilde{\mathbf{s}}_j$
		
\end{algorithm}

\subsection{Phase initialization}
\label{sec:phase_init}

MISI is usually initialized by assigning the mixture's phase to each source. However, its online implementation makes it possible to use an alternative initialization scheme, exploiting phase relationships over time. In particular, we propose here to use a phase model that arise from modeling audio signals as mixtures of sinusoids~\cite{McAuley1986,Krawczyk2014}. It can be shown~\cite{Magron2018} that the phase $\boldsymbol{\varphi}_{j}$ of a source represented as a mixture of slowly-varying sinusoids follows the relationship:
\begin{equation}
\boldsymbol{\varphi}_{j,t}(f) = \boldsymbol{\varphi}_{j,t-1}(f) + 2 \pi l \boldsymbol{\nu}_{j,t}(f),
\label{eq:phase_unwrapping}
\end{equation}
where $\boldsymbol{\nu}_{j,t}(f)$ is the normalized frequency in channel $f$ and time frame $t$.
We propose to use this model as a phase initialization scheme for oMISI (i.e., at line~\ref{algo:phase_init} in Algorithm~\ref{al:oMISI}). The frequencies $\nu$ are estimated from the magnitude spectra using quadratic interpolation around each magnitude peak~\cite{Magron2018,Abe2004}.

\section{Experimental evaluation}
\label{sec:exp}

\subsection{Dataset and protocol}

For evaluation, we consider a single-channel speech separation task. We use the Danish hearing in noise test dataset~\cite{Nielsen2011}. We consider three speaker pairs denoted MF, MM and FF, where M and F stand for male and female respectively, in order to cover all gender combinations. All audio files were recorded with a sampling rate of 44.1 kHz, and down-sampled at 16 kHz in our experiments. The STFT is computed using a $16$ ms long Hann window, $50$ $\%$ overlap, and a zero-padding factor of $2$. The synthesis window is defined as in~\cite{Griffin1984}, so that the STFT is Hermitian, as discussed in Section~\ref{sec:algo_mainup}.

Two scenarios are considered. The Oracle scenario uses the ground truth magnitude spectra of the sources. In the Estim. scenario, they are estimated using the DNN described in~\cite{Naithani2017} (where the interested reader can find details on the DNN architecture and training). This DNN predicts a soft mask that is applied to the mixture to yield magnitude estimates. We compare oMISI to its offline counterpart and to the amplitude mask (AM) used as a baseline~\cite{Naithani2017}.

The separation quality is measured with the scale-invariant signal-to-distortion ratio improvement (SI-SDRi)~\cite{LeRoux2019}. We provide some audio excerpts\footnote{\url{https://magronp.github.io/demos/spl20_omisi.html}} for a subjective evaluation, and the code for reproducing the Oracle scenario experiments \footnote{\url{https://github.com/magronp/omisi}}.

\subsection{Results}

\begin{figure}[t]
    \centering
    \includegraphics[width=.95\columnwidth]{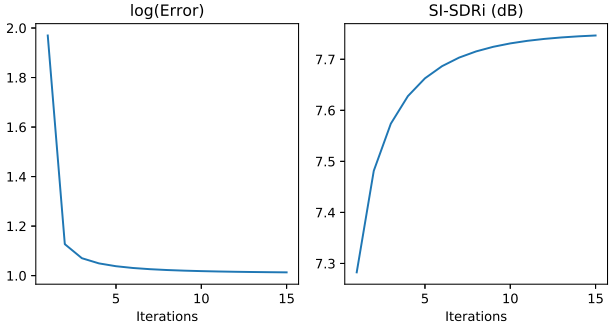}
    \caption{Error~\eqref{eq:mag_function_time} and SI-SDRi over iterations in the Estim. setting.}
    \label{fig:misi_iter}
\end{figure}

We present in Fig.~\ref{fig:misi_iter} the objective loss~\eqref{eq:mag_function_time} and SI-SDRi over iterations for the MISI algorithm using the MF pair in the Estim. setting (similar results are obtained using the other speaker combinations). We observe a non-increasing cost function over iterations, which empirically confirms the convergence of MISI. The SI-SDRi appears to saturate at $15$ iterations, thus we present hereafter the results obtained with this value. Since with oMISI, each frame is processed $(K+1)$ times more as in MISI, we reduce the number of iterations to $15/(K+1)$ for a fair comparison. Finally, note that the SI-SDRi can further increase in the Oracle setting.

\begin{table}[t]
    \caption{Average SI-SDRi in dB (higher is better). Bold fonts correspond to the best performance among online techniques.}
    \vspace{-1em}
	\label{tab:results}
	\scriptsize
	\begin{tabular}{l|c|p{0.55cm}p{0.55cm}|p{0.55cm}p{0.55cm}|p{0.55cm}p{0.55cm}}
    \hline
    \hline
                     &   & \multicolumn{2}{c|}{MF} &  \multicolumn{2}{c|}{MM} &  \multicolumn{2}{c}{FF}  \\
		             & Latency  & Estim. & Oracle & Estim. & Oracle & Estim. & Oracle  \\
		AM      & $16$ ms & $7.5$  & $8.8$  & $5.7$ & $7.3$ & $5.1$ & $7.5$   \\
		MISI         & offline & $7.9$  & $23.8$ & $6.2$ & $22.3$ & $5.4$ & $22.9$   \\
		oMISI  & & & &  & &  &  \\
         \multicolumn{1}{r|}{mix}  & $16$ ms (K=0) & $7.7$  & $16.4$ & $6.1$ & $15.8$ & $5.4$ & $16.9$   \\
         \multicolumn{1}{r|}{mix}  & $24$ ms (K=1) & $7.9$  & $20.2$ & $6.2$ & $19.4$ & $5.4$ & $19.6$   \\
         \multicolumn{1}{r|}{mix}  & $32$ ms (K=2) & $\mathbf{7.9}$  & $\mathbf{21.4}$ & $\mathbf{6.2}$ & $\mathbf{20.4}$ & $\mathbf{5.4}$ & $20.6$   \\
         \multicolumn{1}{r|}{sin}  & $24$ ms (K=1) & $7.8$  & $15.2$ & $6.2$ & $14.6$ & $5.4$ & $\mathbf{20.7}$   \\
       \hline
       \hline
	\end{tabular}
\end{table}

The separation results are reported in Table~\ref{tab:results}. We first remark that MISI improves the performance over the baseline by approximately $0.4$ dB in the Estim. scenario. In the Oracle scenario, the improvement is more significant ($\approx 15$ dB), which highlights the room for improvement for phase recovery. Besides, in the Estim. scenario, we observe that oMISI performs as well as MISI with $K=1$. The performance of oMISI drops slightly for $K=0$, which was expected as no future frame is taken into account: nonetheless, it is still improved compared to AM, and the drop in comparison to the offline method is quite small. Finally, the performance does not further improve for $K=2$ in the Estim. scenario. These results demonstrate the potential of oMISI for real-time applications.

Using one future frame then appears as a good compromise between latency and performance, thus we test the initialization with the sinusoidal phase model with $K=1$. However, this scheme does not improve the performance of oMISI over using the mixture's phase overall. We suggest that the speakers in the MM and MF pairs are sufficiently orthogonal (i.e., less overlapping) in the TF domain, thus the mixture's phase is a good quality initial estimate. Nonetheless, the Estim. results highlight that the FF pair is the most challenging, and the corresponding Oracle results indicate that this initialization scheme has some potential, provided accurate enough magnitude estimates.

\section{Conclusion}
\label{sec:conclusion}

In this paper, we provided the first proof of convergence of the MISI algorithm. We adapted it to operate online without any performance loss, which is an important step towards real-time audio source separation. Future work will focus on deriving alternative spectrogram inversion algorithms based on this theoretical framework, e.g., by replacing the magnitude mismatch distance by a $\beta$-divergence, which is more adapted to audio. This algorithm will also be incorporated into an end-to-end framework for time-domain source separation along with learned and more advanced~\cite{Prusa2016} phase models.

\section{Acknowledgements}

We thank Roland Badeau for his insight on optimization, and Gaurav Naithani for providing the magnitude estimates.

\newpage
\bibliographystyle{IEEEtran}
\bibliography{references}

\begin{thebibliography}{10}
\providecommand{\url}[1]{#1}
\csname url@samestyle\endcsname
\providecommand{\newblock}{\relax}
\providecommand{\bibinfo}[2]{#2}
\providecommand{\BIBentrySTDinterwordspacing}{\spaceskip=0pt\relax}
\providecommand{\BIBentryALTinterwordstretchfactor}{4}
\providecommand{\BIBentryALTinterwordspacing}{\spaceskip=\fontdimen2\font plus
\BIBentryALTinterwordstretchfactor\fontdimen3\font minus
  \fontdimen4\font\relax}
\providecommand{\BIBforeignlanguage}[2]{{%
\expandafter\ifx\csname l@#1\endcsname\relax
\typeout{** WARNING: IEEEtran.bst: No hyphenation pattern has been}%
\typeout{** loaded for the language `#1'. Using the pattern for}%
\typeout{** the default language instead.}%
\else
\language=\csname l@#1\endcsname
\fi
#2}}
\providecommand{\BIBdecl}{\relax}
\BIBdecl

\bibitem{Comon2010}
P.~Comon and C.~Jutten, \emph{Handbook of blind source separation: independent
  component analysis and applications}.\hskip 1em plus 0.5em minus 0.4em\relax
  Academic press, 2010.

\bibitem{Barker2018}
J.~Barker, S.~Watanabe, E.~Vincent, and J.~Trmal, ``The fifth 'chime' speech
  separation and recognition challenge: Dataset, task and baselines,'' in
  \emph{Proc. Interspeech 2018}, September 2018.

\bibitem{Cano2019}
E.~{Cano}, D.~{FitzGerald}, A.~{Liutkus}, M.~D. {Plumbley}, and F.~{Stöter},
  ``Musical source separation: An introduction,'' \emph{IEEE Signal Processing
  Magazine}, vol.~36, no.~1, pp. 31--40, Jan 2019.

\bibitem{Wang2008}
D.~Wang, ``Time-frequency masking for speech separation and its potential for
  hearing aid design,'' \emph{Trends in Amplification}, vol.~12, no.~4, pp.
  332--353, 2008.

\bibitem{Agnew2000}
J.~Agnew and J.~M. Thornton, ``Just noticeable and objectionable group delays
  in digital hearing aids,'' \emph{Journal of the American Academy of
  Audiology}, vol.~11, no.~6, pp. 330--336, June 2000.

\bibitem{Wang2018c}
D.~{Wang} and J.~{Chen}, ``Supervised speech separation based on deep learning:
  An overview,'' \emph{IEEE/ACM Transactions on Audio, Speech, and Language
  Processing}, vol.~26, no.~10, pp. 1702--1726, Oct 2018.

\bibitem{Luo2018}
Y.~{Luo} and N.~{Mesgarani}, ``{TaSNet}: Time-domain audio separation network
  for real-time, single-channel speech separation,'' in \emph{Proc. IEEE
  International Conference on Acoustics, Speech and Signal Processing
  (ICASSP)}, April 2018.

\bibitem{Luo2019}
------, ``{Conv-TasNet}: Surpassing ideal time–frequency magnitude masking
  for speech separation,'' \emph{IEEE/ACM Transactions on Audio, Speech, and
  Language Processing}, vol.~27, no.~8, pp. 1256--1266, August 2019.

\bibitem{Magron2015}
P.~Magron, R.~Badeau, and B.~David, ``{Phase recovery in {NMF} for audio source
  separation: an insightful benchmark},'' in \emph{{Proc. IEEE International
  Conference on Acoustics, Speech and Signal Processing (ICASSP)}}, April 2015.

\bibitem{Gerkmann2015}
T.~Gerkmann, M.~Krawczyk-Becker, and J.~{Le Roux}, ``{Phase Processing for
  Single-Channel Speech Enhancement: History and recent advances},'' \emph{IEEE
  Signal Processing Magazine}, vol.~32, no.~2, pp. 55--66, March 2015.

\bibitem{Mowlaee2016}
P.~Mowlaee, R.~Saeidi, and Y.~Stylianou, ``Advances in phase-aware signal
  processing in speech communication,'' \emph{Speech Communication}, vol.~81,
  pp. 1 -- 29, July 2016, phase-Aware Signal Processing in Speech
  Communication.

\bibitem{Wang2018}
Z.~{Wang}, J.~L. {Roux}, and J.~R. {Hershey}, ``Alternative objective functions
  for deep clustering,'' in \emph{Proc. IEEE International Conference on
  Acoustics, Speech and Signal Processing (ICASSP)}, April 2018.

\bibitem{Magron2018b}
P.~Magron, K.~Drossos, S.~I. Mimilakis, and T.~Virtanen, ``{Reducing
  interference with phase recovery in DNN-based monaural singing voice
  separation},'' in \emph{{Proc. Interspeech}}, September 2018.

\bibitem{Wang2018a}
Z.-Q. Wang, J.~{Le Roux}, D.~Wang, and J.~R. Hershey, ``{End-to-End Speech
  Separation with Unfolded Iterative Phase Reconstruction},'' in \emph{{Proc.
  of Interspeech}}, September 2018.

\bibitem{Wichern2018}
G.~Wichern and J.~{Le Roux}, ``{Phase reconstruction with learned
  time-frequency representations for single-channel speech separation},'' in
  \emph{{Proc. of IWAENC}}, September 2018.

\bibitem{Wang2019a}
Z.~{Wang}, K.~{Tan}, and D.~{Wang}, ``Deep learning based phase reconstruction
  for speaker separation: A trigonometric perspective,'' in \emph{Proc. IEEE
  International Conference on Acoustics, Speech and Signal Processing
  (ICASSP)}, May 2019.

\bibitem{Wisdom2019}
S.~{Wisdom}, J.~R. {Hershey}, K.~{Wilson}, J.~{Thorpe}, M.~{Chinen},
  B.~{Patton}, and R.~A. {Saurous}, ``Differentiable consistency constraints
  for improved deep speech enhancement,'' in \emph{Proc. IEEE International
  Conference on Acoustics, Speech and Signal Processing (ICASSP)}, May 2019.

\bibitem{LeRoux2019a}
J.~{Le Roux}, G.~{Wichern}, S.~{Watanabe}, A.~{Sarroff}, and J.~R. {Hershey},
  ``Phasebook and friends: Leveraging discrete representations for source
  separation,'' \emph{IEEE Journal of Selected Topics in Signal Processing},
  vol.~13, no.~2, pp. 370--382, May 2019.

\bibitem{Gunawan2010}
D.~Gunawan and D.~Sen, ``Iterative phase estimation for the synthesis of
  separated sources from single-channel mixtures,'' \emph{IEEE Signal
  Processing Letters}, vol.~17, no.~5, pp. 421--424, May 2010.

\bibitem{Naithani2017}
G.~{Naithani}, T.~{Barker}, G.~{Parascandolo}, L.~{Bramsl{\o}w}, N.~H.
  {Pontoppidan}, and T.~{Virtanen}, ``Low latency sound source separation using
  convolutional recurrent neural networks,'' in \emph{Proc. IEEE Workshop on
  Applications of Signal Processing to Audio and Acoustics (WASPAA)}, Oct 2017.

\bibitem{Aihara2019}
R.~{Aihara}, T.~{Hanazawa}, Y.~{Okato}, G.~{Wichern}, and J.~L. {Roux},
  ``Teacher-student deep clustering for low-delay single channel speech
  separation,'' in \emph{Proc. 2019 IEEE International Conference on Acoustics,
  Speech and Signal Processing (ICASSP)}, May 2019.

\bibitem{Wang2019}
S.~Wang, G.~Naithani, and T.~Virtanen, ``Low-latency deep clustering for speech
  separation,'' in \emph{Proc. 2019 IEEE International Conference on Acoustics,
  Speech and Signal Processing (ICASSP)}, May 2019.

\bibitem{Mowlaee2015}
P.~Mowlaee and J.~Kulmer, ``{Harmonic phase estimation in single-channel speech
  enhancement using phase decomposition and SNR information},'' \emph{IEEE/ACM
  Transactions on Audio, Speech, and Language Processing}, vol.~23, no.~9, pp.
  1521--1532, September 2015.

\bibitem{Fevotte2011}
C.~F{\'e}votte and J.~Idier, ``{Algorithms for nonnegative matrix factorization
  with the beta-divergence},'' \emph{Neural Computation}, vol.~23, no.~9, pp.
  2421--2456, September 2011.

\bibitem{Magron2018}
P.~Magron, R.~Badeau, and B.~David, ``Model-based {STFT} phase recovery for
  audio source separation,'' \emph{{IEEE/ACM Transactions on Audio, Speech and
  Language Processing}}, vol.~26, no.~6, pp. 1095--1105, June 2018.

\bibitem{LeRoux2013}
J.~{Le Roux} and E.~Vincent, ``Consistent {Wiener} filtering for audio source
  separation,'' \emph{IEEE Signal Processing Letters}, vol.~20, no.~3, pp.
  217--220, March 2013.

\bibitem{Yang2008}
B.~Yang, ``{A study of inverse short-time Fourier transform},'' in \emph{{Proc.
  IEEE International Conference on Acoustics, Speech and Signal Processing
  (ICASSP)}}, April 2008.

\bibitem{Griffin1984}
D.~Griffin and J.~S. Lim, ``{Signal estimation from modified short-time
  {F}ourier transform},'' \emph{IEEE Transactions on Acoustics, Speech and
  Signal Processing}, vol.~32, no.~2, pp. 236--243, April 1984.

\bibitem{Beauregard2005}
G.~T. Beauregard, X.~Zhu, and L.~L. Wyse, ``An efficient algorithm for
  real-time spectrogram inversion,'' in \emph{{Proc. International Conference
  on Digital Audio Effects (DAFx)}}, 2005.

\bibitem{Zhu2006}
X.~{Zhu}, G.~T. {Beauregard}, and L.~{Wyse}, ``Real-time iterative spectrum
  inversion with look-ahead,'' in \emph{Proc. IEEE International Conference on
  Multimedia and Expo}, July 2006.

\bibitem{Zhu2007}
X.~Zhu, G.~T. Beauregard, and L.~L. Wyse, ``Real-time signal estimation from
  modified short-time {Fourier} transform magnitude spectra,'' \emph{IEEE
  Transactions on Audio, Speech, and Language Processing}, vol.~15, no.~5, pp.
  1645--1653, 2007.

\bibitem{McAuley1986}
R.~J. McAuley and T.~F. Quatieri, ``{Speech analysis/Synthesis based on a
  sinusoidal representation},'' \emph{IEEE Transactions on Acoustics, Speech
  and Signal Processing}, vol.~34, no.~4, pp. 744--754, August 1986.

\bibitem{Krawczyk2014}
M.~Krawczyk and T.~Gerkmann, ``{STFT} phase reconstruction in voiced speech for
  an improved single-channel speech enhancement,'' \emph{IEEE/ACM Transactions
  on Audio, Speech, and Language Processing}, vol.~22, no.~12, pp. 1931--1940,
  December 2014.

\bibitem{Abe2004}
M.~Abe and J.~O. Smith, ``{Design criteria for simple sinusoidal parameter
  estimation based on quadratic interpolation of {FFT} magnitude peaks},'' in
  \emph{{Audio Engineering Society Convention 117}}, May 2004.

\bibitem{Nielsen2011}
J.~B. Nielsen and T.~Dau, ``{The Danish hearing in noise test},''
  \emph{International journal of audiology,}, vol.~50, no.~3, pp. 202--8, March
  2011.

\bibitem{LeRoux2019}
J.~L. {Roux}, S.~{Wisdom}, H.~{Erdogan}, and J.~R. {Hershey}, ``{SDR} –
  half-baked or well done?'' in \emph{Proc. IEEE International Conference on
  Acoustics, Speech and Signal Processing (ICASSP)}, May 2019.

\bibitem{Prusa2016}
Z.~Pr\r{u}$\check{\text{s}}$a and P.~L. S{\o}ndergaard, ``Real-time spectrogram
  inversion using phase gradient heap integration,'' in \emph{{Proc.
  International Conference on Digital Audio Effects (DAFx)}}, September 2016.

\end{thebibliography}
\end{document}